\documentclass[%
reprint,
 amsmath,amssymb,
 aps,
]{revtex4-2}

\usepackage[hidelinks]{hyperref}
\hypersetup{
    colorlinks=true,
    allcolors=blue,
    }
    
\usepackage[utf8]{inputenc}
\usepackage[T1]{fontenc}
  \usepackage{subcaption}
\usepackage{comment}
\usepackage{ulem}
\usepackage{colortbl} 
\usepackage{graphicx}% Include figure files
\definecolor{Gray}{gray}{0.9}

\begin{document}

\title{Isotope Shifts in Cadmium as a Sensitive Probe for Physics Beyond the Standard Model}

\author{B. Ohayon}
\email{bohayon@ethz.ch}
\affiliation{Institute for Particle Physics and Astrophysics, ETH Z\"urich, CH-8093 Z\"urich, Switzerland }

\author{S. Hofs\"ass}
\author{J.E. Padilla-Castillo}
\author{S. C. Wright}
\author{G. Meijer}
\author{S. Truppe}
\email{truppe@fhi-berlin.mpg.de. Current address: Centre for Cold Matter, Blackett Laboratory, Imperial College London, Prince Consort Road, London SW7 2AZ, United Kingdom}
\affiliation{Fritz-Haber-Institut der Max-Planck-Gesellschaft, Faradayweg 4-6, 14195 Berlin, Germany}

\author{K. Gibble}
\email{kgibble@psu.edu}
\affiliation{Department of Physics, Pennsylvania State University University Park, PA, 16802 USA}

\author{B. K. Sahoo}
\email{bijaya@prl.res.in}
\affiliation{Atomic and Molecular Physics Division, Physical Research Laboratory, Navrangpura, Ahmedabad 380009, India}

%\keywords{Keyword1, Keyword2, Keyword3}

\begin{abstract}

Isotope shifts (ISs) in atomic energy levels are sensitive probes of nuclear structure and new physics beyond the Standard Model.
We present an analysis of the ISs of the cadmium atom (Cd I) and singly charged cadmium ion (Cd II). ISs of the 229\,nm, 326\,nm, 361\,nm and 480\,nm lines of Cd I are measured with a variety of techniques; buffer-gas-cooled beam spectroscopy, capturing atoms in a magneto-optic-trap, and optical pumping.
IS constants for the D$_1$ and D$_2$ lines of Cd II are calculated with high accuracy by employing analytical response relativistic coupled-cluster theory in the singles, doubles and triples approximations.
Combining the calculations for Cd II with experiments, we infer IS constants for all low-lying transitions in Cd I.
We benchmark these constants as calculated via different many-body methods.
Our calculations for Cd II enable nuclear charge radii of Cd isotopes to be extracted with unprecedented accuracy. 
The combination of our precise calculations and measurements shows that King Plots for Cd~I can improve the state-of-the-art sensitivity to a new heavy boson by up to two orders of magnitude.

\end{abstract}

%\flushbottom
\maketitle

\section{Introduction}

\begin{figure*}[!tb]
%\centering
\includegraphics[width=0.9\linewidth]{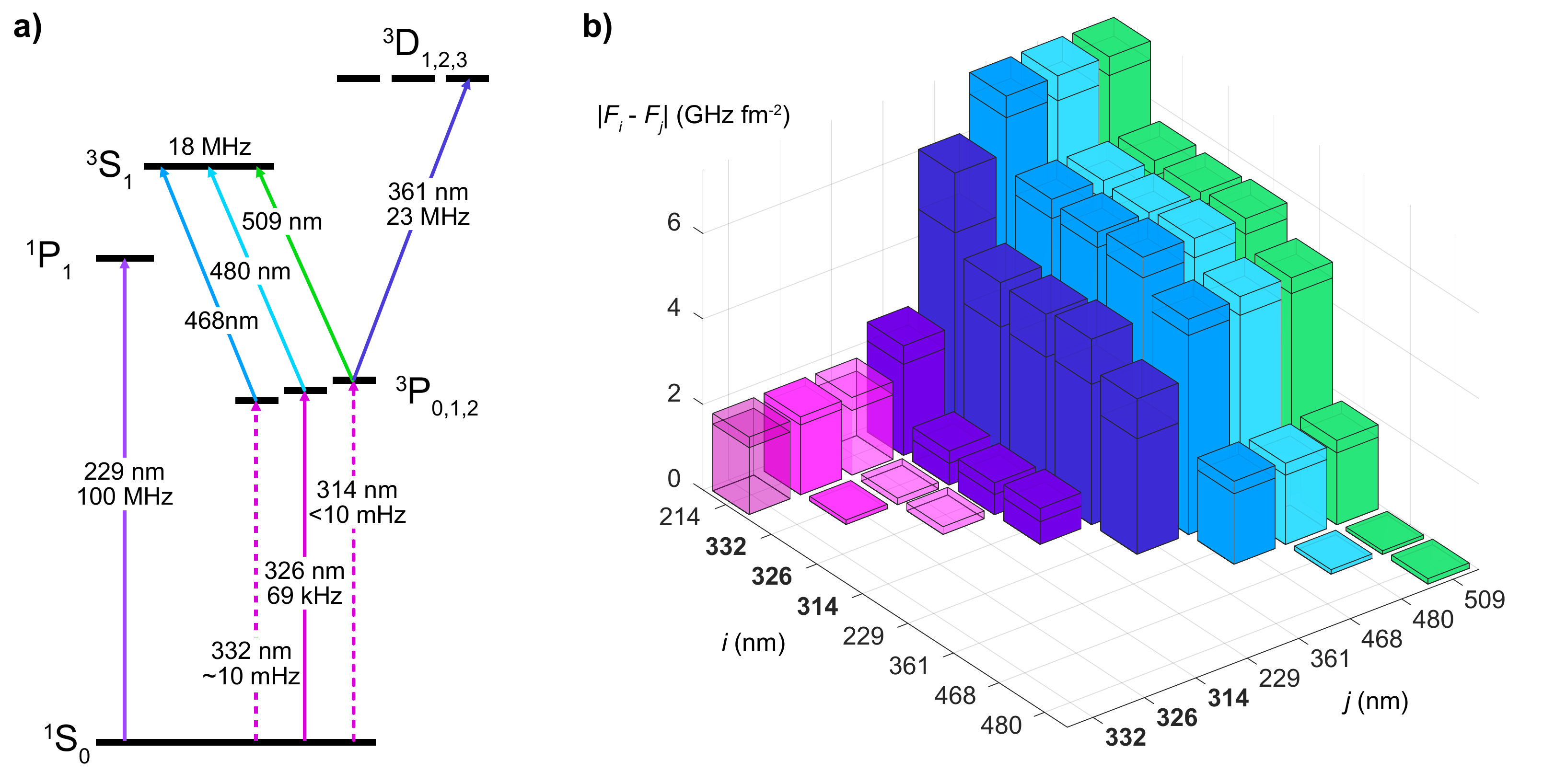}
\caption{a: Low-lying energy levels of Cd I. The isotope shifts of the indicated transitions are discussed in this work. Broken lines represent transitions whose ISs are deduced from combinations of the ISs of measured ISs. 
b: Magnitudes of the difference of field shift constants. The top boxes corresponds to the $69\%$ confidence interval. The transitions are ordered for clarity, with the narrow transitions in bold, and the 214 nm transition is for Cd II.}
\label{fig:Levels}
\end{figure*}

The most accurately determined quantities are transition frequencies of optical clocks \cite{2021-OC}, which are measured to better than $10^{-18}$.
This superb spectroscopic precision enables stringent searches for new physics \cite{2014-DM, 2018-TableTop, 2020-GR, 2021-var}.
One approach compares the isotope shifts (IS) of two or more transitions as in a King Plot (KP) \cite{1963-King}, in which a deviation from a linear behaviour \cite{2017-KPnonlin,2017-KPnon,2018-Neut,2020-GKP} can be a signature for beyond Standard Model (BSM) physics, such as a new boson mediating a force between electrons and neutrons.
King Plot searches are applicable to systems possessing narrow optical transitions for which ISs can be measured with high precision, even with  
extraordinary sub-Hz accuracy using common-mode noise rejection \cite{2019-Sr, 2022-DiffClock}.
Searches for deviations from linearity with KPs require a minimum of four stable (or long-lived) even-even isotopes, which severely restricts the number of candidate elements. 

BSM searches through KP non-linearity were performed for two elements in different regimes.
On the low atomic number ($Z$) side are measurements in calcium ($Z=20$); the lightest element with more than three stable even-even isotopes. A KP comprised of two transitions with different relativistic effects for Ca II, with a characteristic precision of $20\,$Hz, shows no non-linearity, translating directly to limits on several BSM scenarios \cite{2020-Ca}.
As experimental precision improves, the potential for new physics searches using KPs in light elements such as Ca is expected to be limited by the difficulty of calculating the standard model (SM) contribution to the KP non-linearity resulting from high-order recoil effects in a light many-electron system \cite{2021-Li,2021-NP}.

In the high $Z$ regime, the effects of a new massive boson are more pronounced \cite{2017-KPnonlin,2021-NP}, while recoil effects are heavily suppressed.
However, other sources of SM non-linearity become dominant making the interpretation of an observed non-linearity more involved.
For Yb ($Z=70$), ISs in several ionic and atomic lines were measured with high-precision \cite{2020-Yb,2021-Yb, 2021-Yb2, 2022-Yb}. 
A recent analysis encompassing all of the experimental data indicates a strong non-linearity attributed to nuclear deformation \cite{2022-Yb}, which currently  cannot be calculated {\it ab initio} with sufficient precision.
This behaviour is particularly pronounced and follows from Yb isotopes being amongst the most deformed in the stable region of the nuclear chart \cite{2017-Deform}. 
A non-linearity from an additional source was identified, whose origin is currently being studied \cite{2021-Yb,2022-Yb}. 
Considering the cases above, once a non-linearity is observed, it is not straightforward to interpret it as a signature of new physics. Pronounced non-linearities in several systems will add clarity.
Moreover, there is a trade-off between lower values of $Z$, where both the non-linearities arising from the Standard Model (SM) and those that come from new physics effects are less pronounced; and higher values of $Z$, where several sources of non-linearity may be difficult to disentangle due theoretical intractability.

Cadmium ($Z=48$) sits in between these two limits.
Having a relatively simple atomic structure and narrow transitions \cite{2018-326MOT, 2019-NarrowLine}, it is a prime candidate for new physics searches via KP non-linearities.
Being close to the $Z=50$ proton shell closure makes Cd nuclei much less deformed than e.g. Yb, suppressing a major source of SM non-linearity.
Cd possesses six stable even-even isotopes, which could allow up to three different sources of non-linearity to be identified. In this work, we present measurements of ISs in the neutral cadmium atom (Cd I). We combine these measurements with literature values and highly accurate calculations of IS constants for transitions in the singly charged cadmium ion (Cd II). This enables us to map out all of the IS constants of the low-lying transitions in Cd I shown
in Fig. \ref{fig:Levels}, and identify the promising combinations for BSM searches. We compare the obtained IS constants with recent calculations and discuss the current state of the art in calculating these constants for two-valence systems.
Such calculations are needed to assess the sensitivity of KPs to specific new physics models.
We also combine calculated IS constants with measurements in a long chain of short-lived isotopes to yield highly accurate charge radii differences, which are needed to determine the nuclear contributions to KP non-linearities \cite{2021-Radii, 2021-Yb}.
Finally, we discuss the prospects for searches for BSM physics via KPs in Cd in light of our analysis. 

\section{Theory}

The IS $\delta \nu_i^{A,A'}$ between isotopes with mass numbers $A,A'$, at the precision of the measurements analyzed here, can be written as \cite{2013-King}
\begin{equation}
\delta \nu^{A,A'}_i = 
F_i  \delta \lambda^{A,A'} + K_i\mu^{A,A'}.
\label{IS}
\end{equation}
Here $\mu^{A,A'}=1/M_{A}-1/M_{A'}$ the inverse nuclear mass difference, $i$ denotes a particular atomic transition, and $\delta \lambda^{A,A'}=\Sigma_{k} S_{2k} (\delta r_c^{2k})^{A,A'}$ is the nuclear parameter \cite{1987-Blun}, which is expanded in a series of
even charge moments
\begin{equation}
    r_c^k=\frac{\int r^k \rho(r)  r^2 dr}{\int \rho(r)  r^2 dr}.
    \label{moments}
\end{equation}
For brevity, we denote the root mean square (RMS) charge radius $\sqrt{r_c^2}$ here as $r_c$. 
The constants $F_i$ and $K_i$ are the field shift and mass shift, which depend only on the transition in question at this level.
To calculate these constants, the nuclear charge distribution can be approximated using a Fermi distribution
\begin{equation}
    \rho(r)=\frac{\rho_0}{1+e^{(r-b)/a}},
    \label{2pF}
\end{equation}
where $\rho_0$ is a normalization constant, and $b$ and $a$ are determined from electron scattering experiments \cite{1975-Gi, 1976-Li}.

The operator for the field shift is given by
\begin{equation}
F=
-\sum_e \frac{\delta V_{nuc}(r_c, r_e)}{\delta \lambda}
\approx - \sum_e \frac{\delta V_{nuc}(r_c, r_e)}{\delta \langle r_c^2 \rangle}(1+f_\lambda)
\end{equation}
where the electrostatic potential $V_{nuc}$ is a function of $r_c$, the radial distance of an electron is $r_e$, and the sum is over all electrons of the system.
The contribution from higher nuclear moments $f_\lambda$ is estimated as
\begin{equation}
f_\lambda \approx 
-S_4\frac{r_c^4}{ r_c^2}-
 S_6\frac{ r_c^6}{ r_c^2}=2.8(3)\%,
 \label{Blund}
\end{equation}
where the Seltzer coefficients $S_4$ and $S_6$ are estimated for the $5S$ levels of Cd II \cite{1987-Blun}, and the ratio of moments is taken from two parameterizations of electron scattering experiments \cite{1975-Gi, 1976-Li}. The uncertainty of $f_\lambda$ is estimated from the model-dependence of the charge distribution of Eq. \ref{2pF}, and its variation between isotopes.

The mass shift constant can further be split into a normal mass shift (NMS) and specific mass shift (SMS).
The operators to determine the NMS and SMS constants are defined in a relativistic framework as \cite{1987_Palmer, 1998-Shab}
{\scriptsize
\begin{eqnarray}
K^{\text{NMS}} &=& \frac{1}{2}\sum_i \left ({\vec p}_i^{~2} - \frac{\alpha Z}{r_i} {\vec \alpha}_i^D \cdot {\vec p}_i  - \frac{\alpha Z}{r_i} ({\vec \alpha}_i^D \cdot {\vec C}_i^1) {\vec C}_i^1 \cdot {\vec p}_i \right )
\label{eq:NMS}
\end{eqnarray}}
and
{\scriptsize
\begin{eqnarray}
K^{\text{SMS}} &=& \frac{1}{2} \sum_{i\ne j} \left ({\vec p}_i \cdot {\vec p}_j - \frac{\alpha Z}{r_i} {\vec \alpha}_i^D \cdot {\vec p}_j  - \frac{\alpha Z}{r_i} ({\vec \alpha}_i^D \cdot {\vec C}_i^1) ({\vec p}_j \cdot {\vec C}_j^1) \right ),
\label{eq:SMS}
\end{eqnarray}
}
where the $p_i$'s are the components of the momentum operator, $\alpha^D$ is the Dirac matrix, and $C^1$ is the Racah operator. 
We note that there are slight modifications in the definitions of the $F$, $K^{\text{NMS}}$ and $K^{\text{SMS}}$ operators if we include quantum electrodynamics (QED) interactions. These contributions are expected to be smaller than our reported uncertainties.

A linear relation results after applying Eq. \ref{IS} to the measured ISs of two transitions $(i,j)$:
\begin{equation}
    \delta \bar{\nu}_i^{A,A'}=F_{ij}\delta\bar{\nu}_j^{A,A'}+K_{ij}
    \label{KP}
\end{equation}
with the modified ISs $\delta \bar \nu_i^{A,A'}=\delta \nu_i^{A,A'}/\mu^{A,A'}$, the slope $F_{ij}=F_i/F_j$, and the offset $K_{ij}=K_i-F_{ij}K_j$.
In this work we find that the linear relation of Eq. \ref{KP} holds 
at the $\sim \,$MHz-level for all Cd transition pairs.
It may therefore be used to project IS constants from one transition to another.
We calculate IS constants for transitions in Cd II using the analytical response (AR) relativistic coupled-cluster (RCC) method, and project them using Eq. \ref{KP} to atomic transitions.

\section{The AR-RCC method}

We begin with the Dirac-Coulomb (DC) Hamiltonian:

{\scriptsize
\begin{eqnarray}\label{eq:DC}
H_{DC} &=& \sum_i \left [c\mbox{\boldmath${\vec \alpha}$}_i^D \cdot {\vec p}_i+(\beta_i-1)c^2+V_{nuc}(r_i)\right] +\sum_{i,j>i}\frac{1}{r_{ij}}, 
\end{eqnarray}}

where $c$ is the speed of light, $\mbox{\boldmath${\vec \alpha}$}^D$ and $\beta$ are the Dirac matrices, ${\vec p}$ is the single particle momentum operator.
$\sum_{i,j}\frac{1}{r_{ij}}$ represents the Coulomb potential between the electrons located at the $i^{th}$ and $j^{th}$ positions. Corrections from the Breit and QED interactions are estimated by adding the corresponding potential terms as
in Ref. \cite{2021-LiCC}, yielding the atomic Hamiltonian $H_0$.

We consider the nuclear charge distribution of Eq. (\ref{2pF}) to define the potential $V_{nuc}$ \cite{1985-Est} 

\begin{eqnarray}
 V_{nuc}(r) = -\frac{Z}{\mathcal{N}r} \times \ \ \ \ \ \ \ \ \ \ \ \ \ \ \ \ \ \ \ \ \ \ \ \ \ \ \ \ \ \ \ \ \ \ \ \ \ \ \ \  \nonumber\\
 {\scriptsize
\left\{\begin{array}{rl}
\frac{1}{b}(\frac{3}{2}+\frac{a^2\pi^2}{2b^2}-\frac{r^2}{2b^2}+\frac{3a^2}{b^2}P_2^+\frac{6a^3}{b^2r}(S_3-P_3^+)) & \mbox{for $r_i \leq b$}\\
\frac{1}{r_i}(1+\frac{a62\pi^2}{b^2}-\frac{3a^2r}{b^3}P_2^-+\frac{6a^3}{b^3}(S_3-P_3^-))                           & \mbox{for $r_i >b$} ,
\end{array}\right.  
}
\label{eq12}
\end{eqnarray}

with
\begin{eqnarray}
\mathcal{N} &=& 1+ \frac{a^2\pi^2}{b^2} + \frac{6a^3}{b^3}S_3,  \nonumber \\
 \ \ \ \ S_k &=& \sum_{l=1}^{\infty} \frac{(-1)^{l-1}}{l^k}e^{-lb/a}, \ \ \  \nonumber \\
 \ \ \ \ P_k^{\pm} &=& \sum_{l=1}^{\infty} \frac{(-1)^{l-1}}{l^k}e^{\pm l(r-b)/a} . 
\end{eqnarray}
We determine these constants for the ground state, $[4d^{10}] 5s$, and the first two excited states, $[4d^{10}]5p_{1/2}$ and $[4d^{10}]5p_{3/2}$ of Cd II to study the ISs of its D$_1$ and D$_2$ lines.

In the RCC theory {\it ansatz}, the wave function of the above atomic states are constructed as \cite{1982-Lind,1989-Muk,2004-Pb} 
\begin{eqnarray}
 |\Psi_v \rangle = e^T \{1+S_v \} |\Phi_v \rangle,
 \label{eqcc}
\end{eqnarray}
where $|\Phi_v \rangle$ is a mean-field wave function from 
a Dirac-Hartree-Fock (DHF) treatment, and $T$ and $S_v$ are the excitation operators that account for electron correlations from the core orbitals and valence orbital, respectively.
The subscript $v$ denotes the valence orbital associated with the respective state. It is introduced to uniquely identify the states having the common closed-shell configuration $[4d^{10}]$.
Considering the IS operators, $F$, $K^{\text{NMS}}$ and $K^{\text{SMS}}$, denoted as $O$ in the total Hamiltonian as $H=H_0 + \eta O$, we express the above wave function 
in the AR-RCC formulation as
\begin{eqnarray}
 |\Psi_v \rangle = |\Psi_v^{(0)} \rangle + \eta  |\Psi_v^{(1)}\rangle + {\cal O}(\eta^{2}),
\end{eqnarray}
with an energy 
\begin{eqnarray}
 E_v = E_v^{(0)} + \eta E_v^{(1)} + {\cal O}(\eta^{2}).
\end{eqnarray}
Here $\eta$ is nominally equal to one and is introduced for the perturbation expansion in orders of $O$, denoted by the superscripts.
$E_v^{(0)}$ corresponds to the contribution from $H_0$ and $E_v^{(1)}$ includes first-order contributions from $O$ and electron correlations. The above can be implemented in the RCC theory by expanding the RCC operators as
\begin{eqnarray}
 T= T^{(0)} + \eta T^{(1)}+ {\cal O}(\eta^2) 
 \end{eqnarray}
 and
 \begin{eqnarray}
 S_{v}= S_{v}^{(0)} + \eta S_v^{(1)}+ {\cal O}(\eta^2) .
\end{eqnarray}
The ${\cal O}(\eta^2)$ contributions are usually small and neglected in IS calculations. 
However, these non-linear contributions can be significant when
probing BSM physics. The zeroth-order RCC operator amplitudes follow from 
\begin{eqnarray}
&& \langle \Phi_0^* |  \bar{H}_0 | \Phi_0 \rangle = 0  \label{eqamp0} \\
\text{and} && \nonumber \\
 && \langle \Phi_v^* | \{ (\bar{H}_0-E_v^{(0)}) S_v^{(0)} \} + \bar{H}_0 | \Phi_v \rangle = 0, \label{eqamp}
\end{eqnarray}
where $\langle \Phi_{0,v}^* |$ denotes for all possible excited Slater determinants and $\bar{H}_0= \left ( H_0 e^{T^{(0)}} \right )_{conn}$, 
where the subscript $conn$ denotes for only connected terms in the expansion. With these amplitudes of the RCC operators, we calculate the zeroth-order energies as
\begin{eqnarray}
&& E_0^{(0)} = \langle \Phi_0 | \bar{H}_0 | \Phi_0 \rangle \label{eqeng0} \\
\text{and} && \nonumber \\
&& E_v^{(0)} = \langle \Phi_v | \bar{H}_0 \{ 1+ S_v^{(0)} \} | \Phi_v \rangle, \label{eqeng}
\end{eqnarray}
where $E_0^{(0)}$ is the energy of the common closed-core of the considered atomic states, $[4d^{10}]$ for Cd II. In the actual calculations, we consider a normal-ordered Hamiltonian defined with respect to $[4d^{10}]$. Therefore, $E_v^{(0)}$ corresponds to the electron affinity, $E_v^{(0)}-E_0^{(0)}$, of the valence orbital relative to the $[4d^{10}]$ configuration. 
We note that Eqs. (\ref{eqamp}) and (\ref{eqeng}) are coupled.

In the AR-RCC theory, we calculate the desired FS, NMS and SMS constants as the first-order energy corrections $E_v^{(1)}\equiv \langle \Psi_v^{(0)}|O|\Psi_v^{(0)}\rangle$ using the following expression \cite{2020-In,2021-LiCC}
\begin{eqnarray}
E_v^{(1)} = \langle \Phi_v | \overline{H_0} S_v^{^{(1)}}  + (\overline{H_0} T^{(1)} + \overline{O}) \{ 1 + S_v^{^{(0)}} \} | \Phi_v \rangle ,  \label{eqeng1}
\end{eqnarray}
where $\overline{O}=(Oe^{T^{(0)}})_{conn}$. 
Here, the normal-ordered form of operators again yield calculated values relative to the contributions from the $[4d^{10}]$ configuration. 
The amplitudes of the first-order perturbed RCC operators are given by
\begin{eqnarray}
 && \langle \Phi_0^* |  \overline{H}_0 T^{(1)} +\overline{O} | \Phi_0 \rangle = 0. \label{eqamp0} \\
 \text{and} && \nonumber \\
 && \langle \Phi_v^* | (\overline{H}_0-E_v^{(0)}) S_v^{(1)} + \nonumber \\
 && \left (\overline{H}_0 T^{(1)} +\overline{O} \right ) \{1+ S_v^{(0)} \}| \Phi_v \rangle = 0. \label{eqamp1}
\end{eqnarray}
We use Gaussian type orbitals (GTOs) \cite{1951-Boys} to construct the single particle DHF wave functions. The large and small radial components of the DHF orbitals, $P(r)$ and $Q(r)$, are expressed using these GTOs as
\begin{eqnarray}
 P(r) = \sum_{k=1}^{N_k} c_k^{\cal L} \zeta_{\cal L} r^l e^{-\eta_0 \gamma^k r^2}
\end{eqnarray}
and 
\begin{eqnarray}
 Q(r) = \sum_{k=1}^{N_k} c_k^{\cal S} \zeta_{\cal L} \zeta_{\cal S} \left (\frac{d}{dr} + \frac{\kappa}{r}\right ) r^l e^{-\eta_0 \gamma^k r^2} ,
\end{eqnarray}
where $l$ is the orbital quantum number, $\kappa$ is the relativistic angular momentum quantum number, $c_k^{{\cal L}({\cal S})}$ are the expansion coefficients, $\zeta_{{\cal L}({\cal S})}$ are the normalization factors of GTOs, $\eta_0$ and $\gamma$ are optimized GTO parameters for a given orbital, and $N_k$ represents the number of GTOs. To construct the GTOs, we use $\eta_0=  0.00715, 0.0057, 0.0072, 0.0052$ and $0.0072$ for $s$, $p$, $d$, $f$ and $g$ orbitals, respectively, with corresponding $\gamma$ values $1.92, 2.04, 1.97, 2.07$ and $2.54$. Since our orbitals are not bounded, we integrate to an upper radial limit of $r=500$ a.u. on a grid using a 10-point Newton-Cotes Gaussian quadrature method.
Our numerical calculations use exponential grids with a step-size of 0.016 a.u. and 1200 grid points, and the coefficients are determined by the Roothan equation in the relativistic framework.
To reduce the required computational resources, we have limited the virtual space by considering all possible single and double excitations in the AR-RCC theory (AR-RCCSD method) for the 1$-$20$s$, 2$-$20$p$, 3$-$20$d$, 4$-$17$f$, and 5$-$14$g$ orbitals. The AR-RCCSDT method adds triple excitations to the above single and double excitations only for the 16$s$, 16$p$, 14$d$, 10$f$, and 5$g$ orbitals.

\section{Calculated IS constants for Cd II}

\begin{table*}
\centering
\caption{\label{tab:res} Cd II Energies and IS constants for the ground and first two excited states, and their transitions for the D$_1$ ($226.5\,$nm) and D$_2$ ($214.4\,$nm) lines.
}
   \begin{subtable}{0.8\textwidth}
        \centering
        \caption{\label{tab:transition}Electron affinities and transition energies   
        (\textrm{cm}$^{-1}$)}
\begin{ruledtabular}
\begin{tabular*}{\textwidth}{@{\extracolsep{\fill}}llllll}

        Method & ~ ~ $E_{5S}$ &  ~ ~ $E_{5P1/2}$& ~ ~ $E_{5P3/2}$&  ~ ~ $E_{\textrm{D1}}$ &  ~ ~ $E_{\textrm{D2}}$ \\
        \hline
 
  DHF      & 124568& 84903  &  82871& 39666 & 41698  \\
 RCCSD& 136013& 91858   &89357& 44155 & 46657  \\
 RCCSDT & 136717& 92401& 98974& 44316 & 46843 \\
~$+\Delta$Breit & ~ ~ $-83$ & ~ $-83$ & ~ $-50$ & ~ ~ ~  0. & ~ $-33$  \\
~$+\Delta$QED  &~~ $-100$ & ~ ~ $-4$&~ ~ $-6$ &~ $-97$ &~ $-94$ \\
SUM & 136533(200)& 92314(150)&89818(150)& 44218(97) & 46715(104)  \\
Exp \cite{1949-SP,1956-BA} & 136374.7(1) & 92238.7(1)&89756.2(1) & 44136.1(1)    & 46618.6(1)   \\

\end{tabular*}
\end{ruledtabular}
\end{subtable}

\vspace*{2mm}

\begin{subtable}{0.8\textwidth}
\caption{\label{tab:normal}Normal mass shift  (GHz u) }
\centering
\begin{ruledtabular}
\begin{tabular*}{\textwidth}{@{\extracolsep{\fill}}lllllll}

Method & $K_{\textrm{NMS},5S}$& $K_{\textrm{NMS},5P1/2}$& $K_{\textrm{NMS},5P3/2}$& $K_{\textrm{NMS,D1}}$& $K_{\textrm{NMS,D2}}$\\
\hline
 DHF         & 5952&3374&3163&2578 & 2789  \\
 AR-RCCSD    &2113&1381&1339&~ 732 &~ 774  \\
 AR-RCCSDT  &  2204(20)&1470(15)&1426(15)&~ 734(7)  &~ 778(8) \\
~$+\Delta$Breit &~ $-4$&~ $-4$&~ $-2$&~ ~~ 0. &~~ $-2$  \\
~$+\Delta$QED  & ~ $-7(2)$ &~ ~  0. &~ ~  0. &~~ $-7(2)$ &~~ $-7(2)$  \\
SUM &  2194(20)& 1466(15)&1424(15)&~ 727(8) &~ 769(8)  \\
 Scaling &  2243&1517 &1476  &~ 726    & ~ 767     \\
\end{tabular*}
\end{ruledtabular}

\vspace*{2mm}

%\begin{subtable}{0.8\textwidth}
\caption{\label{tab:specific}Specific Mass Shift  (GHz u) }
\centering
\begin{ruledtabular}
\begin{tabular*}{\textwidth}{@{\extracolsep{\fill}}llllll}

Method & $K_{\textrm{SMS},5S}$& $K_{\textrm{SMS},5P1/2}$& $K_{\textrm{SMS},5P3/2}$& $K_{\textrm{SMS,D1}}$& $K_{\textrm{SMS,D2}}$\\
\hline
 DHF         & $-2775$&$-1753$&$-1488$& $-1022$ & $-1287$  \\
 AR-RCCSD    & ~~ 1288 & ~ ~~ 123 & ~ ~~ 256 & ~~ 1165 & ~~ 1032  \\
 AR-RCCSDT  & ~~ 1343(15)&~ ~~ 129(5)&~ ~~ 260(7)&~~ 1214(16) &~~   1083(17) \\
~$+\Delta$Breit &~ ~ ~  12& ~ ~ ~ ~ 5& ~ ~ ~ ~ 4& ~ ~ ~~ 7 &~ ~ ~~ 8  \\
~$+\Delta$QED  & ~ ~~ ~ 4(1)& ~ ~ ~  $-2(0.)$& ~ ~ ~ $-2(0.)$& ~ ~ ~~  6(1) & ~ ~ ~~  5(1)  \\
%     \hline 
 SUM &  ~~ 1359(15)& ~ ~~ 132(5)& ~ ~~ 263(7)& ~~ 1226(16) & ~~  1096(17) \\

\end{tabular*}
\end{ruledtabular}
\end{subtable}

\vspace*{2mm}

\begin{subtable}{0.8\textwidth}
\caption{\label{tab:total}Total Mass Shift  (GHz u) }
\centering
\begin{ruledtabular}
\begin{tabular*}{\textwidth}{@{\extracolsep{\fill}}lccccc}

Method & $K_{5S}$& $K_{5P1/2}$& $K_{5P3/2}$& $K_{\textrm{D1}}$& $K_{\textrm{D2}}$\\
 \hline
$K_\textrm{SMS}$+$K_\textrm{NMS}$  & 3552(25)&1598(16)&1686(17)& 1954(18) ~ & 1866(19) ~  \\
CI-MBPT \cite{2021-Julian}& &&&1770(300) & 1667(300) \\
CKP \cite{2021-Julian}&  &&&& 2199(507) \\

\end{tabular*}
\end{ruledtabular}
\end{subtable}

\vspace*{2mm}

\begin{subtable}{0.8\textwidth}
\caption{\label{tab:field}Field Shift (MHz\,fm$^{-2}$) }
\centering
\begin{ruledtabular}
\begin{tabular*}{\textwidth}{@{\extracolsep{\fill}}llllll}

Method & ~ ~ $F_{5S}$&  ~ ~ $F_{5P1/2}$&  ~ ~ $F_{5P3/2}$&  ~ ~ $F_{\textrm{D1}}$&  ~ ~ $F_{\textrm{D2}}$\\
\hline
 DHF         & $-4778$& ~  $-59$& ~ ~ $-0.$& $-4719$ & $-4778$  \\
 AR-RCCSD    &  $-6140$& $-177$&$-100$&$-5963$ & $-6040$ \\
 AR-RCCSDT   & $-6227(20)$&$-232(5)$&$-152(5)$&$-5995(21)$ & $-6075(21)$\\
~$+\Delta$Breit & ~ ~  ~~  22&  ~ ~  ~~ 1&  ~~ ~ ~ 1&  ~ ~  ~~ 21 &  ~ ~~ ~  21  \\
 ~$+\Delta$QED  & ~ ~~ 147(37)& ~ ~ ~~ 4(1)& ~ ~ ~~ 3(1)& ~ ~~ 143(36) & ~ ~~ 144(36)   \\
 ~-$f_\lambda$ &  ~  $-172(17)$ & ~ ~~ ~ 0& ~ ~~ ~ 0& ~ $-172(17)$&  ~ $-172(17)$  \\
  SUM & $-6230(45)$&$-277(5)$&$-149(5)$& $-6003(45)$ & $-6082(45)$   \\
  CI-MBPT \cite{2021-Julian}&&&& $-6067(300)$ & $-6144(300)$ \\
CKP \cite{2021-Julian}& &&& & $-6621(530)$ \\

\end{tabular*}
\end{ruledtabular}
\end{subtable}
\end{table*}

\begin{table*}[htb]
\centering
\caption{\label{tab:IS}
Isotope shifts of Cd I in MHz, relative to  $^{114}$Cd.
Derived values are in italics; for the $332\,$nm clock line, from this work and previous measurements of $468\,$nm ISs \cite{1981-WTM}, and similarly for the narrow $314\,$nm transition using $509\,$nm ISs \cite{2018-ISOLDE}.}

\begin{ruledtabular}
\begin{tabular}{llllllll}
A  &  $229\,$nm & $326\,$nm &   $480\,$nm & $361\,$nm &  $\mathit{332}\,$nm & $\mathit{314}\,$nm \\
\hline
%\rowcolor{Gray}  
106  & ~$1818.1(3.5)^{a}$  & 1911.2(3.3)$^{a}$ & $-798.5(1.0)^c$ & $-607.6(3.0)^b$ & $\mathit{1911.4(4.7)}$ & $\mathit{1915.5(3.9)}$ \\
%\rowcolor{Gray}  
 &  ~$1748(11)$ \cite{2021-TSB} & 1913.0(1.0)$^{b}$ &&&& \\
%\rowcolor{Gray}  
 &  & $1921(26)$ \cite{1959-KT,1987-KBB326} &&&& \\
 
108  & ~$1336.5(3.4)^a$  & $1399.4(3.3)^{a}$ & $-586.7(1.0)^c$ &$-447.4(3.0)^b$  & $\mathit{1404.5(3.7)}$ & $\mathit{1403.9(3.0)}$\\
 & ~$1258(9)$ \cite{2021-TSB}   & $1402.4(1.0)^{b}$ &&&& \\
  &  &  $1410(40)$ \cite{1987-KBB326}   \\
%\rowcolor{Gray}  
110  & ~~ $865.0(3.3)^a$   & $909.3(3.3)^{a}$  & $-383.4(1.0)^c$ &  $-293.9(1.0)^b$  & $\mathit{914.1(2.1)}$ & $\mathit{913.7(2.1)}$\\
%\rowcolor{Gray}  
  & ~~ $826(6)$ \cite{2021-TSB}  &  $914.7(1.0)^{b}$  &&&& \\
%\rowcolor{Gray}  
  & ~~ $906(35)$ \cite{1961-KT} &  909(13) \cite{1950-BS,1959-KT,1964-Les}  &&&& \\

112  & ~~ $407.5(3.3)^a$   &  $426.3(3.3)^{a}$  &  $-183.1(1.0)^c$ & $-142.2(1.0)^b$ & $\mathit{428.9(1.7)}$ & $\mathit{429.7(1.6)}$\\
 &  ~~ $392(5)$ \cite{2021-TSB}   & $429.9(1.0)^{b}$ &&&&   \\
 & ~~ $396(30)$ \cite{1961-KT}  &  $403(11)$ \cite{1950-BS,1959-KT,1964-Les} &&&&   \\

%\rowcolor{Gray}  
116  & $-316.1(3.3)^a$  & $-326.9(3.3)^{a}$ & ~~  $152.7(1.0)^c$ & ~~ $122.0(1.0)^b$& $\mathit{-320.6(1.7)}$ & $\mathit{-320.2(1.7)}$\\
%\rowcolor{Gray}  
&   $-299(4)$ \cite{2021-TSB}     & $-321.5(1.0)^{b}$ &&&&  \\
%\rowcolor{Gray}  
&  &  $-279(12)$ \cite{1959-KT,1964-Les} &&&&  \\

%\multicolumn{7}{l}{$^{a}$ Beam measurement, $^{b}$ MOT edge, $^{c}$ MOT repumping}
\end{tabular}
$^{a}$ Beam measurement, $^{b}$ Blue MOT edge, $^{c}$ MOT optical pumping rate
\end{ruledtabular}
\end{table*}

An important test of \textit{ab initio} methods to obtain accurate atomic wave functions is to compare the calculated energies with the measured ones.
In Table \ref{tab:transition} we present the calculated electron affinities of the ground and first two excited states of Cd II. From these values, we determined the excitation energies of the D$_1$ and D$_2$ transitions given in Table \ref{tab:transition}.
The RCCSDT results show that triple excitations contribute more to the electron affinities than to the excitations energies, and that the Breit and QED corrections are significant.
The uncertainties of the calculated energies are 
estimated from the convergence.
We reproduce the experimental transition energies to $0.2\%$, an order of magnitude of improvement over CI-MBPT calculations for these transitions \cite{2021-Julian}.
The high accuracy with which experimental energies are reproduced using the RCC method validates the accuracy of our wave functions for the following IS constants calculation.
However, accurate wavefunctions are a necessary but insufficient condition for obtaining accurate IS constants. Various many-body methods for calculating the operators produce significantly different results for the IS constants; the AR method performs well against stringent benchmarks \cite{2021-LiCC, 2022-Na}.

We present IS constants using the DHF, AR-RCCSD and AR-RCCSDT methods in Table \ref{tab:res}.
The AR-RCCSDT uncertainty follows from the numerical convergence as well as a perturbative estimation of partial quadrupole excitations.
An uncertainty of $25\%$ is assigned to the QED corrections and $10\%$ to the higher moment corrections.
%
%% NMS discussion %%
Focusing first on the NMS constants, they can be obtained in the nonrelativistic limit by invoking the Virial theorem \cite{1928-Virial}. This results in the scaling law $K_{\textrm{NMS}}\approx E\cdot m_e$, where $E$ is the experimental energy and $m_e$ the electron mass in atomic units.
However, for a medium mass system such as Cd II, it is not \textit{a prioi} clear how applicable this method is.
Table \ref{tab:normal} gives our results for $K_{\textrm{NMS}}$ calculated using the corresponding relativistic operator in Eq. \ref{eq:NMS}.
We see that for the D$_1$ and D$_2$ transitions, the scaling law agrees with our calculation within the $1\%$ numerical accuracy. This behaviour is attributed to strong cancellations of relativistic contributions to the NMS in transitions with the same principal quantum number \cite{2012-Scaling}.
For the ground state the scaling law is accurate to $2-3\%$. 
Triple excitations contribute significantly to the NMS constants for the electron affinities, even though these quantities are evaluated with a one-body operator.
We thus expect that for similar transitions in lighter systems, for which the contribution of electron correlations to the NMS is more pronounced, using the scaling-law could be preferable to a fully relativistic calculation.

%% SMS discussion %%
The specific mass shift is associated with the two-body operator of Eq. \ref{eq:SMS} and thus strongly affected by high-order electron correlations, which are challenging to estimate \cite{1983-Na}.
Even in modern calculations, an uncertainty of $10-20\%$ in $K_\textrm{SMS}$ is usually given (see e.g. \cite{2016-Mn,2020-In,2021-Al,2021-K, 2021-Julian}). 
Our results in Table \ref{tab:specific} are quoted with an a accuracy of $1-2\%$, attributed to higher-order electron correlations, which we estimate perturbatively.
This is in contrast to $K_\textrm{NMS}$ and the energies, whose 
uncertainties follow from numerical convergence.
We see that the sign of $K_\textrm{SMS}$ changes between the DHF and AR-RCCSD calculations. This points to the importance of strong electron correlations in $K_\textrm{SMS}$, with a $10\%$ difference coming from triple excitations.
Owing to the two-body nature of the SMS operator, triple excitations in the AR-RCC method take several months to calculate at a typical high-performance computing facility. 
The total mass shift constant $K=K_\textrm{NMS}+K_\textrm{SMS}$ in Table \ref{tab:total} agrees well with, and is an order of magnitude more precise than, a recent CI-MBPT calculation for both ionic transitions \cite{2021-Julian}, as well as a calibrated King plot (CKP) estimation combining muonic x-ray and electron scattering measurements with precise isotope shifts for the D$_2$ transition \cite{2021-Julian}.
The contributions of the Breit and QED interaction to $K$ are found to be negligible at the current level of precision.

%% FS discussion %%
The calculation of the field shift operator is considered robust for a number of systems. Various methods typically agree at the few percent level \cite{2005-Saf,2020-Yb,2018-CaCI, 2001-Saf,2003-TelFS,2003-Julian,2021-Ca,2021-LiCC, 2022-Na}. 
In Cd II, $F$ convergences quickly, with triple excitations contributing only $0.6\%$.
This yields an $0.35\%$ uncertainty of $F$ at the AR-RCCSDT level, which could be reduced further with more computational resources.
However, we find that QED effects, which are often
not taken into account in such calculations, are not negligible. In fact, our final uncertainty for
$F$ is dominated by the systematic uncertainty associated with QED correction only being present at the Hamiltonian level. This is expected to be even more pronounced in heavier systems such as Yb, where calculations of $F$ are given with sub-percent numerical precision \cite{2021-SMYb}.
Our results in Table \ref{tab:field} agree with the CI-MBPT and a CKP results of  \cite{2021-Julian}.
To our knowledge, this is the first calculation of $F$ with sub-percent accuracy (computational and systematic) for a system with more than $19$ electrons.

\section{IS measurements in the Cd atom}

\begin{figure*}[htbp]
\centering
\includegraphics[width=\linewidth]{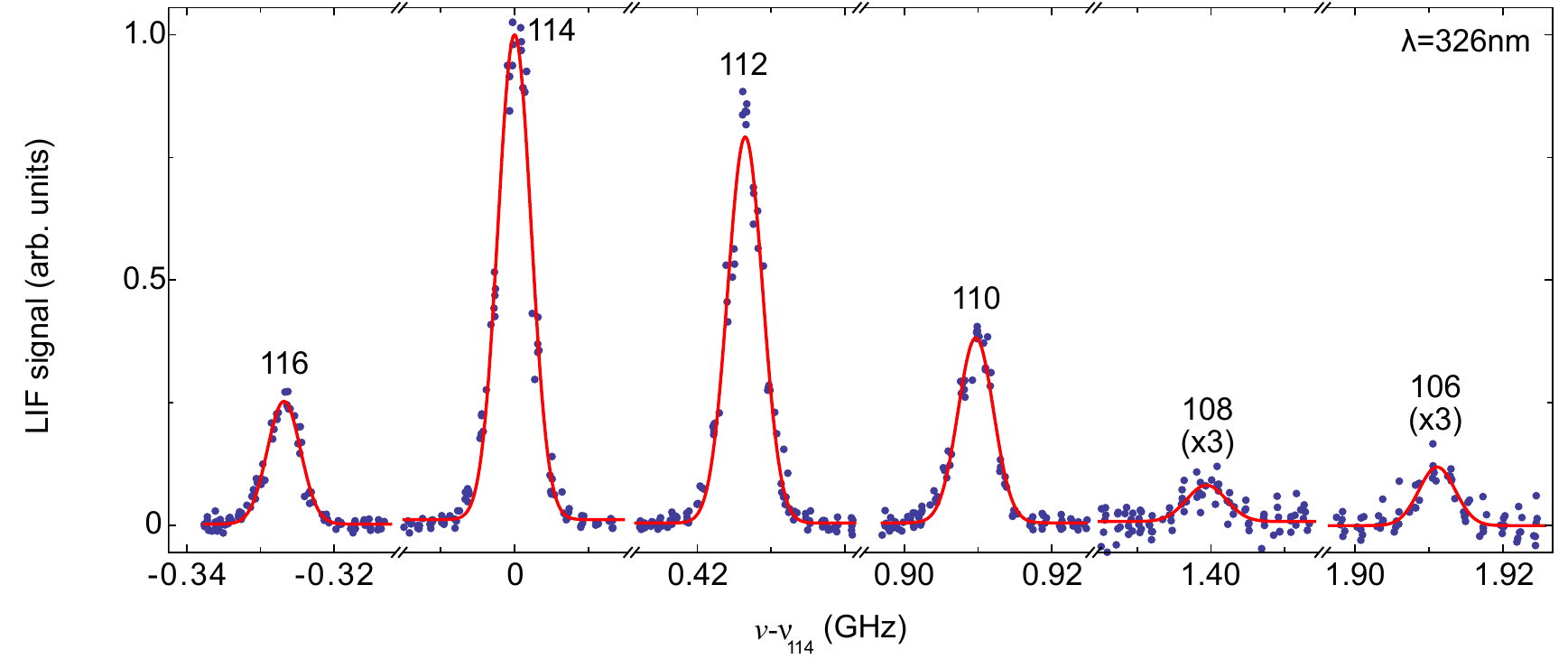}
\caption{
Isotope shifts of the Cd I bosons at 326~nm,
measured with a cryogenic helium-cooled atomic beam and a frequency-doubled dye laser.
The fermionic hyperfine transitions are beyond this spectral range.
}
\label{fig:beam326}
\end{figure*}

In this section we describe our IS measurements of four transitions of Cd I for all stable bosonic isotopes. 
Measurements are done with two experimental systems employing different frequency calibration procedures. A more detailed description of the experimental systems, as well as results for the fermionic isotopes and absolute wavenumbers, will be given in upcoming publications \cite{Hofsaess2022,2022-KG}.
Our results are given in Table \ref{tab:IS}, along with available previous measurements, which disagree significantly in several cases.
The two independent experimental results at $326$\,nm agree to within $0.5-1.6$ combined standard errors, whereas the disagreements with these previous measurements are as large as four combined errors.

\subsection{Atomic Beam Measurements of 229 nm and 326 nm ISs}

We carried out laser induced fluorescence spectroscopy with a buffer-gas-cooled beam. A pulsed beam is produced by laser ablation of a Cd target, which is mounted in a buffer gas cell. Cryogenic helium gas (3\,K) flows continuously through the cell, cools the hot atoms and extracts them into an atomic beam with a mean forward velocity of roughly 120\,m/s. The laser light intersects the atomic beam perpendicularly, and fluorescence is collected on a UV-sensitive photomultiplier tube. The transverse velocity of the atoms is reduced to below about 0.5\,m/s by a slit directly before they enter the fluorescence detector. The resulting Doppler broadenings are 1.5\,MHz (326\,nm) and 2.1\,MHz (229\,nm).

\begin{figure*}[tb]
\centering
\includegraphics[width=\linewidth]{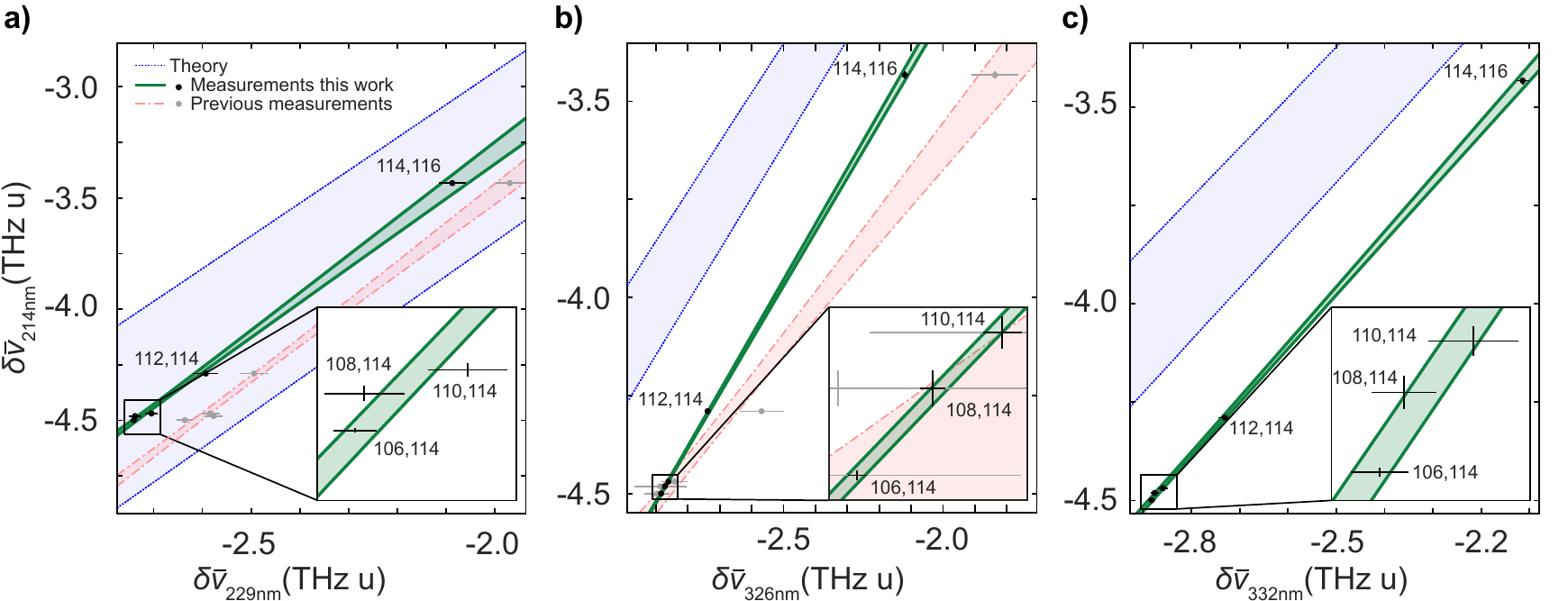}
\caption{King Plots of selected Cd transitions.
The vertical axis is the modified isotope shift of the $214\,$nm line from \cite{2021-Julian}, and the AR-RCCSDT theoretical prediction from Table \ref{tab:res}. The data for the horizontal axes includes the experimental measurements from this work in Table \ref{tab:IS}, the theory prediction for $229\,$nm is from CI-MBPT \cite{2021-Julian}, and for $326$ and $332\,$nm from MCDHF \cite{2021-SM}.
Fits to previous experimental data are given for the $229\,$nm \cite{2021-TSB} and $326\,$nm lines \cite{1950-BS, 1959-KT, 1964-Les, 1987-BDK326, 1987-KBB326}.
The bands correspond to 68\% confidence intervals.}
\label{fig:KP}
\end{figure*}

Due to the large natural linewidth of the 229\,nm line, $\Gamma/(2\pi)=100(3)\textrm{MHz}$ \cite{Hofsaess2022}, a high-precision IS measurement is hampered by the overlap of the different isotopes in the spectrum. To improve the accuracy, we use enriched targets to precisely determine the lineshape. A Lorentzian function fits the data well, suggesting that Doppler broadening is indeed negligible. We then measure the IS of the $(110,112)$ pair, for which the overlap in the spectrum is most severe. This is done by taking multiple spectra while alternating two separate enriched ablation targets for $^{110}$Cd and $^{112}$Cd. This IS is then fixed in a fit to the spectrum recorded using an ablation target with natural abundance. To resolve the frequencies of the fermionic isotopes in the spectrum, we vary their amplitude relative to the bosonic isotopes by changing the polarisation angle of the excitation laser with respect to the detector axis. The weighted means for each isotope are averaged and the final result is given in Table \ref{tab:IS} and plotted in Fig. \ref{fig:KP}a.
Our results are more accurate and differ considerably from those of a recent measurement \cite{2021-TSB}. This discrepancy manifests itself as a horizontal offset in the fits to the experimental data portrayed in Fig. \ref{fig:KP}a.

The $326\,$nm intercombination line of Cd has a $69\,$kHz natural linewidth that enables both cooling to low temperatures and precise spectroscopy. To measure the ISs of this line in an atomic beam, we use a frequency-doubled dye laser (Sirah Matisse 2DX with a Spectra Physics Wavetrain doubling module), whose linewidth is stabilised to better than 100\,kHz short term and 1\,MHz shot-to-shot stability. A typical spectrum is shown in Fig. \ref{fig:beam326}, where the larger Doppler broadened linewidth of 5.6(2)~MHz (Gaussian FWHM) arises from inclusion of atoms with higher velocity in the spectrum. The isotope shifts of five independent measurements have an average standard deviation of 0.8\,MHz, consistent with the wavemeter (HighFinesse WS8-10) resolution of 0.4\,MHz at the fundamental wavelength. This statistical uncertainty is small compared with the systematic uncertainty in the wavemeter measurement of the laser frequency. To place an upper bound on this systematic uncertainty, we measure nearby optical transitions of atomic copper in the same beam machine. The D$_1$ (327.5\,nm) and D$_2$ (324.8 nm) lines lie roughly 1.3\,nm to either side of the 326\,nm line in Cd. %The hyperfine splittings of the ground state has been precisely measured with an accuracy of about 10\,kHz \cite{Ting1957}.
We reproduce the precisely measured ground state hyperfine intervals of $^{65}$Cu (12568.81(1)~MHz) and $^{63}$Cu (11733.83(1)\,MHz) \cite{Ting1957} with a mean deviation of 3.1$\pm$2.0~MHz for $^{63}$Cu and 1.8$\pm$3.7~MHz for $^{65}$Cu. By measuring the wavemeter linearity with a temperature and pressure stabilized Zerodur cavity we find a systematic uncertainty of 3.3~MHz. We assign this as our systematic frequency uncertainty for the 326~nm and 229~nm spectra of Cd I.

\subsection{Measurements of 326, 480 and 361 nm ISs with magneto-optic traps}

We determine the ISs of the 326 line by measuring the sharp blue-edge of the fluorescence \cite{1993-Xe} from a $326\,$nm magneto-optic trap (MOT) \cite{2022-KG}. The blue edge, where the cooling and trapping breaks down, is of order $20\,$kHz wide ($90\%$ to $10\%$) and serves as a precise reference for ISs. We directly load a $326\,$nm MOT of the abundant isotopes, $^{110}$Cd to $^{116}$Cd, from an effusive $1.2\,$cm diameter source of natural Cd that is $2.2\,$cm from the MOT. To capture the atoms, we use $60$ to $150\,$mW of $326\,$nm light that is frequency modulated with a peak-to peak amplitude of $8.6\,$MHz \cite{2019-NarrowLine,2022-KG}. The three retro-reflected MOT beams have a diameter of $8.5\,$mm. A lower-intensity cooling stage follows the MOT loading with no applied magnetic field gradient or frequency modulation, during which fluorescence is measured.

We produce $326\,$nm light with doubly-resonant sum frequency generation (SFG) of $542\,$nm and $820\,$nm light in a Beta Barium Borate crystal \cite{2018-326MOT}. The 542 nm light is produced by frequency doubling a fiber-amplified $1083\,$nm extended cavity diode laser (ECDL). An $820\,$nm ECDL and a tapered amplifier supply the $820\,$nm light. 
Isotope shifts are measured relative to various modes of a temperature-tuned, 4-mirror, rectangular reference cavity. Its free-spectral range is $905.03\,$MHz, its tangential and sagittal transverse mode splittings are $201.04\,$MHz and $140.84\,$MHz, and the cavity linewidth is $350\,$kHz. The $1083\,$nm ECDL is locked to the 00, 01 and 10 modes with an adjustable frequency offset provided by a double-passed acousto-optic modulator (AOM).
The frequency of the $820\,$nm ECDL is measured relative to resonances of the same reference cavity \cite{2018-326MOT}, by slowly frequency modulating a double-passed AOM with a zero-to-peak amplitude of $1.2\,$MHz or less. Within 1 to 5 minutes after measuring the $326\,$nm frequency offset for each isotope, the absolute frequency of a $1083\,$nm cavity resonance is checked relative to a molecular iodine line via saturated absorption at $542\,$nm.  We correct for the cavity drift between the measurements for each isotope, which is typically of order $1\,$MHz, and had a maximum drift of $4.2\,$MHz.
The measurement uncertainty is given by the less than $1\,$MHz centering precision of the $820\,$nm light on a cavity resonance.

For $^{106}$Cd and $^{108}$Cd, with $1.25\%$ and $0.89\%$ natural abundances, we enhance the $326\,$nm MOT loading rate by capturing atoms on the $23\,$MHz wide $361\,$nm $^{3}P_{2}$-$^{3}D_{3}$ transition, after optical pumping from $^{3}P_{1}$ and $^{3}P_{0}$ to $^{3}P_{2}$ via the $^{3}S_{1}$ state (see Fig. \ref{fig:Levels}a). The $361\,$nm light is generated via doubly-resonant SFG of $1083\,$nm and $542\,$nm light, producing $70$ to $200\,$mW \cite{2018-326MOT,2022-KG}. To load the $326\,$nm MOT, the $361\,$nm MOT is inhibited by turning off the $480\,$nm $^{3}P_{1}$ - $^{3}S_{1}$ optical pumping light and allowing the atoms to equilibrate in the $326\,$nm MOT for $57\,$ms before the start of the $326\,$nm cooling and detection phase. 
Our results for the ISs for the $326\,$nm line are given in Table \ref{tab:IS} and Fig. \ref{fig:KP}b.
They are more accurate than our measurement of the same line with a beam, and comprise an order of magnitude of improvement compared with the weighted average of previous measurements. A disagreement of 4 combined errors is found for the (114,116) pair, previously measured with interferometry of fluorescence \cite{1959-KT,1964-Les}. 
This disagreement manifests as different slopes of the King Plot portrayed in Fig. \ref{fig:KP}b.

To measure ISs for the $480\,$nm transition, we use all four laser sources to enhance the MOT loading and then inhibit the $480\,$nm light for $28\,$ms to transfer atoms to the $326\,$nm MOT and cool them. We then turn on a $4.4\,$ms low intensity pulse of $480\,$nm light and observe the resulting $361\,$nm fluorescence. 
We generate the $468\,$nm and $480\,$nm optical pumping light using single-pass SFG of $1083\,$nm light and $823\,$nm and  $862\,$nm light in fiber-coupled PPLN waveguides. The $468\,$nm and $480\,$nm light are combined on a beamsplitter and the two output beams illuminate the atoms on three nearly orthogonal axes before being retro-reflected. The $862\,$nm laser, and thereby the frequency of the $480\,$nm light, is slowly stabilized with a wavemeter and is monitored with the same reference cavity, using a number of transverse modes and a zero-to-peak frequency modulation of 1.2 MHz via a double-passed AOM.
We determine the center frequency of the $17.5\,$MHz wide transition with a resolution of $0.8\,$MHz using a slow square wave frequency modulation of the $480\,$nm light by $\pm 4\,$MHz with a double-passed AOM. 
Our $480\,$nm ISs in Table \ref{tab:IS} are the first measurements for this transition. By combining them with our $326\,$nm intercombination line ISs along with previous measurements for $468\,$nm \cite{1981-WTM}, we determine the ISs for the $332\,$nm clock transition (see Fig. \ref{fig:KP}c), and similarly for the narrow $314\,$nm transition using measured $509\,$nm ISs \cite{2018-ISOLDE}.

Finally, we measure the ISs of the the $361\,$nm $^{3}P_{2}$-$^{3}D_{3}$ transition. We again use the sharp blue-edge of the cooling as the reference for the ISs \cite{1993-Xe}. Here, we use the reference cavity stability to measure frequency offsets relative to $542\,$nm iodine saturated absorption resonances  \cite{2002-I2}. The blue edge is determined to better than $0.4\,$MHz and we estimate that the blue edges, especially for the slower loading $^{106}$Cd and $^{108}$Cd MOTs, may have systematic errors of order $3
\,$MHz, much less than the $23\,$MHz $361\,$nm transition linewidth.

\section{Benchmarking IS constants calculations} 

\begin{table}[!tbp]
\centering
\caption{IS constants of atomic transitions deduced from our calculations (Table \ref{tab:res}) projected using King Plots with experimental data from Table \ref{tab:IS} and the literature \cite{2021-TSB,1959-KT,1987-KBB326,1964-Les,2018-ISOLDE,2021-Julian}.
}
\begin{ruledtabular}
\begin{tabular}{llll}
i (nm)      & $F_i$ (MHz$\,$fm$^{-2}$)& $K_i$ (GHz u) & Method \\
\hline
%\rowcolor{Gray}  
$229$  & $-3764(134)$ & ~$ 1199(104)$ &  \\
%\rowcolor{Gray}  
         & $-4024(200)$ & ~$1428(220)$ & CI-MBPT \cite{2021-Julian}  \\

$332$  & $-4391(86)$ & ~$1712(43)$ & \\
         & $-4660(160)$ & ~$1629(19)$ &MCDHF \cite{2021-SM} \\
         
%\rowcolor{Gray}  
$326$ & $-4354(62)$ & ~$1673(43)$ & \\
%\rowcolor{Gray}  
& $-4680(160)$ & ~$1616(17)$ & MCDHF \cite{2021-SM} \\
%\rowcolor{Gray}  
& $-4559(230)$ & ~$1865(400)$ & CI-MBPT \cite{2021-Julian}  \\
%\rowcolor{Gray}  
& $-4420(340)$ & ~$1717(330)$ & CKP \cite{2004-FrickeCd}  \\

$314$  & $-4421(86)$ & ~$1739(61)$ & \\
%\rowcolor{Gray}  
$468$ &~~ $1150(49)$ & ~~ $-1(47)$ & \\

$480$ & ~~ $1122(40)$ & ~ ~ $30(41)$ & \\
%$441.6$ & $15040(299)$ & $-6066(218)$ &  \\\\
%\rowcolor{Gray}  
$509$ & ~~ $1180(19)$ & ~~$-32(24)$ & \\
%\rowcolor{Gray}  
        & ~~ $1228(60)$ & ~~$-63(400)$ & CI-MBPT \cite{2021-Julian}  \\
        
        $361$  & ~ $-654(42)$ & ~$-237(59)$ & \\

\end{tabular}
\end{ruledtabular}
\label{tab:Atom}
\end{table}

The constants of eq. \ref{KP} for the D$_1$ and D$_2$ ionic transitions can be extracted from our calculation (Table \ref{tab:res}) with high accuracy because the uncertainties for each transition are highly correlated.
We find $F_{D2,D1}=1.0131(2)$ and $K_{D2,D1}=-114(2)\,$GHz\,u.
A Monte-Carlo fit to the available experimental data \cite{2018-ISOLDE,1964-Les, 1969-CK, 1971-BK, 1976-BSS} yields $F_{D2,D1}=1.02(6)$ and $K_{D2,D1}=-43(229)\,$GHz\,u, which agrees with our calculation, within the large experimental uncertainty. 
To better test our calculation, recent ion-trap IS measurements for the D$_2$ transition \cite{2021-Julian} could be extended to the D$_1$ transition.

To test calculation in Cd I, we transform our constants for the Cd II D$_2$ transition to transitions in Cd I using eq. \ref{KP}, fitting experimental ISs from this work and the literature \cite{2021-TSB,1959-KT,1987-KBB326,1964-Les,2018-ISOLDE,2021-Julian}. 
Here, we do not consider fermionic isotopes, for which eq. \ref{KP} may not be a good approximation at the few MHz level \cite{2021-SM}. 
The resulting IS constants are given in Table \ref{tab:Atom}. Their uncertainty is dominated by that of the measured isotope shifts of the atomic lines.
The IS constants of tables \ref{tab:res} and \ref{tab:Atom} agree with, and are up to an order of magnitude more precise than, the CI-MBPT calculations \cite{2021-Julian} for both $F$ and $K$ for all tested ionic and atomic transitions.
We thus validate both the central values and the uncertainty estimation of CI-MBPT, which is applicable for calculations of both linear and non-linear IS constants in a variety of atomic systems.

\begin{table*}[!htbp]
\centering
\caption{Updated differences of charge radii of even-even Cd isotopes in fm$^2$. 
This work and \cite{2021-Julian} use measurements for the D2 line from  \cite{2018-ISOLDE,2021-Julian}, while the radii extracted using atomic factors from the MCDHF method \cite{2021-SM} utilize measurements in the intercombination line summarized in \cite{2004-FrickeCd}.
The charge radius is obtained by adding $(r_c)^{114}=4.614(3)\,$fm in quadrature.}
\begin{ruledtabular}
\begin{tabular}{clllll}

A & ~ $(\delta r^2)^{A,114}$ &~   CKP \cite{2021-Julian} & MCDHF \cite{2021-SM} & $(\delta r^2)^{A+2,A}$ &  Ref \cite{2004-FrickeCd}\\
\hline
100 & $-1.466(12)$  & $  -1.409(19) $ &            & 0.2876(23) &  \\
102 & $-1.178(9) $  & $ -1.135(14) $ &            & 0.2480(20) & 0.28(12) \\
104 & $-0.930(8) $  & $  -0.897(10)$ &            & 0.2149(18) & 0.235(95) \\
106 & $-0.714(6) $  & $ -0.690(8)  $ & $-0.662(24)$& 0.1899(15) & 0.182(16) \\
108 & $-0.524(4) $  & $  -0.506(6) $ & $-0.478(19)$ & 0.1817(15) & 0.183(15) \\
110 & $-0.343(3) $  & $ -0.331(4) $ &  $-0.310(12)$& 0.1791(14) & 0.174(13) \\
112 & $-0.164(1) $  & $  -0.158(2) $ & $-0.145(6) $ & 0.1635(13) & 0.157(10) \\
114 & ~~ $0$ &   ~~ 0   & ~~ 0           & 0.1360(11) & 0.129(10) \\
116 & ~~ $0.136(1)$  & ~~ 0.134(2)  & ~~ 0.115(5)   & 0.1108(27) & 0.083(22) \\
118 & ~~ $0.248(2)$  & ~~ 0.244(6)  &               & 0.1003(25) & 0.084(50) \\
120 & ~~ $0.349(3)$  & ~~ 0.344(8)  &               & 0.0892(19) &  \\
122 & ~~ $0.439(4)$  & ~~ 0.434(11) &               & 0.0825(31) &  \\
124 & ~~ $0.520(5)$  & ~~ 0.514(14) &               & 0.0760(9) &  \\
126 & ~~ $0.596(5)$  & ~~ 0.590(18) &               & 0.0755(9) &  \\
128 & ~~ $0.671(6)$  & ~~ 0.666(21) &               & 0.0443(8) &  \\
130 & ~~ $0.716(7)$  & ~~ 0.713(24) &               &          &  \\

\end{tabular}
\end{ruledtabular}
\label{Tab:Rad}
\end{table*}

For the intercombination line, our measurements produce a significantly larger KP slope than prior results, increasing from $F_{214,326}=0.98(8)$ to $F_{214,326}=1.397(17)$, corresponding to the green solid and red dashed lines in figure \ref{KP}b. Combining the slope with $F_{214}=-6082(45)\,$MHz\,fm$^{-2}$ from Table \ref{tab:field} shifts $F_{326}$ from $-6180(478)\,$MHz\,fm$^{-2}$ to $-4354(62)\,$MHz\,fm$^{-2}$, closer to the value obtained for the $229\,$nm transition which is expected as they share an s-state.
Our $F_{326}$ and $K_{326}$ constants agree with and are more accurate than CI-MBPT calculations \cite{2021-Julian} and an empirical CKP determination \cite{2004-FrickeCd}. 
However, Fig. \ref{fig:KP}b shows a slight disagreement with a recent calculation via MCDHF \cite{2021-SM}.
As summarized in Table \ref{tab:field}, the deviation for $F_{326}$ is two combined standard errors and for $K_{326}$ by one.
A similar deviation is apparent for the clock transition at $332\,$nm (Fig. \ref{fig:KP}c).
The changes in $F_{326}$ and $K_{326}$ and the updated intercombination line ISs account for the $20-50\%$ differences between the ISs calculated in  \cite{2021-SM} and those summarized in Ref.  \cite{2004-FrickeCd}.
Whereas the range of values of $F_i$ is a few percent, we note that the many-body QED corrections are estimated to contribute 3$\%$. Further improvements in field shift calculations should include these QED effects, especially for heavier elements.

\section{Improved Charge radii}

\begin{figure*}[tb]
\centering
\includegraphics[width=\linewidth]{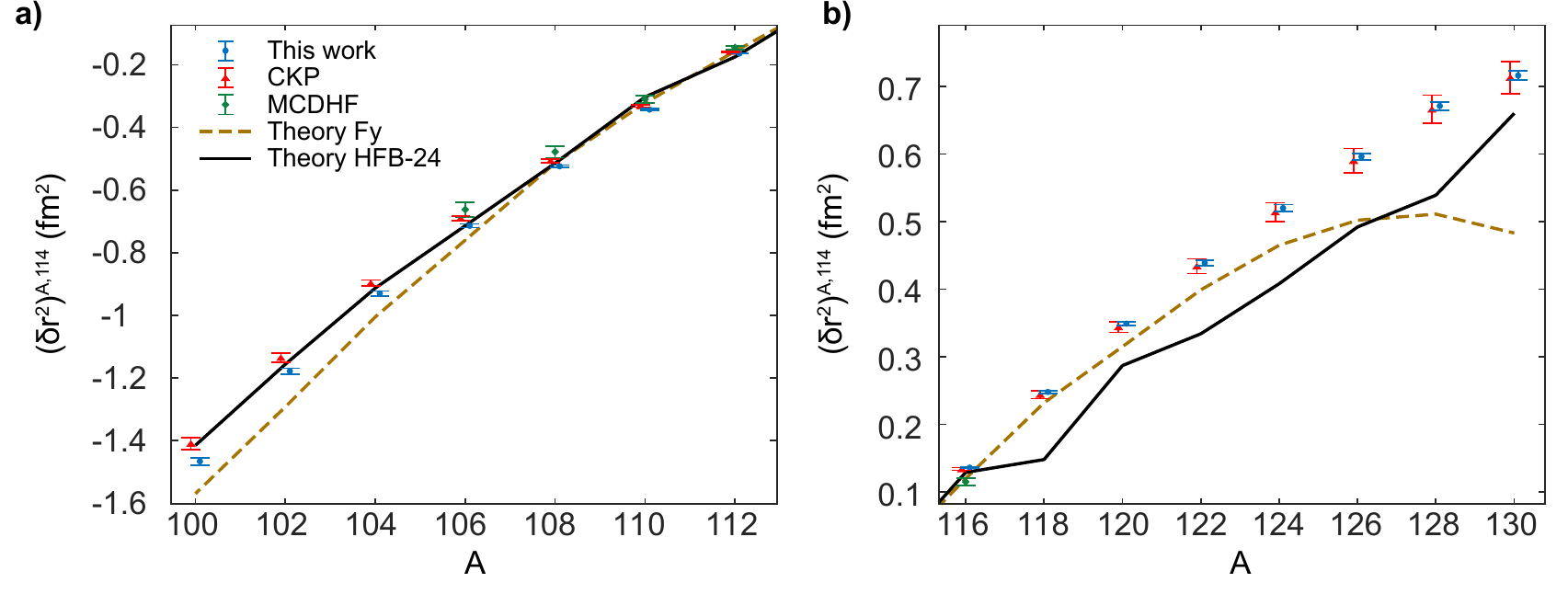}
\caption{
Squared RMS charge radii of even-even Cd isotopes relative to $^{114}$Cd. a: proton rich, b: neutron rich.
The data points correspond to columns $2-4$ of Table \ref{Tab:Rad}. The nuclear theory lines are from the Fayans functional (Fy), extended by information in the radii of Ca \cite{2018-ISOLDE}, and Skyrme-Hartree-Fock-Bogoliubov mass formula (HFB-24) \cite{2021-Julian}.
}
\label{fig:Rad}
\end{figure*}
High-precision optical isotope and isomer shifts for the Cadmium chain were measured at ISOLDE using collinear laser spectroscopy \cite{2016-Isomer, 2018-ISOLDE}.
To extract charge radii differences from spectroscopic data, eq. \ref{IS} is often calibrated with the radii of stable isotopes extracted from a combination of muonic x-ray energies and electron scattering data \cite{2004-FrickeCd}.
This commonly used CKP method has several disadvantages:
it is currently applicable only to elements possessing at least three stable or long-lived isotopes;
the determined IS constants are limited by the knowledge of nuclear polarization corrections to muonic levels, as well as their unknown correlations;
the calibration coefficients from electron scattering experiments are in many cases unknown (e.g. for $^{106,108,111,113}$Cd) or nuclear-model-dependent (e.g. for $^{110,112,114,116}$Cd); and the resulting IS constants $K$ and $F$ are highly correlated, leading to diverging uncertainties of extracted radii outside of the calibrated region.

The advent of high-accuracy many-body calculations of IS constants allows differences of charge radii to be extracted directly from optical measurements with no input, beyond small higher-moment corrections, from muonic x-ray or electron scattering experiments.
In Table \ref{Tab:Rad} and Fig. \ref{fig:Rad} we compare $(\delta r^2)^{A,114}=r^2_{A}-r^2_{114}$ extracted from experimental data  \cite{2018-ISOLDE,2021-Julian} and our calculated $F_{214}$ and $K_{214}$ to those determined recently using a CKP method \cite{2021-Julian}.
The radii of odd isotopes and isomers \cite{2016-Isomer} could also be extracted after calculating off-diagonal hyperfine elements.
For $^{118,120,122}$Cd, measured only using an atomic transition at $509\,$nm, we use $F_{509}$ and $K_{509}$ given in Table~\ref{tab:Atom}.
Our reported uncertainty is $0.7-0.9\%$, dominated by the calculated $F_{214}$ uncertainty, in turn stemming from a systematic uncertainty in the QED correction. For all $(\delta r^2)^{A,114}=r^2_{A}-r^2_{114}$, it is smaller than the uncertainty from the CKP method, especially for neutron-rich isotopes, for which the uncertainty can be as much as a factor of 4 smaller.
A slight disagreement is found for mass numbers 112 and lower as seen in Fig. \ref{fig:Rad}a, which we ascribe to the limitations the CKP method listed above.
Our improved charge radii increase the disagreement with state-of-the-art nuclear theory calculations \cite{2018-ISOLDE, 2021-Julian} for the most neutron rich isotopes (Fig. \ref{fig:Rad}b), while improving the agreement for those that are proton-rich (Fig. \ref{fig:Rad}a).

The benefit of extracting radii directly from calculated atomic constants is even more pronounced for the ladder differences $(\delta r^2)^{A+2,A}=r^2_{A+2}-r^2_{A}$, which are highly sensitive to the uncertainties of muonic x-ray measurements. Such differences are useful, e.g. for identifying quantum phase transitions in nuclear shapes \cite{2012-QPT,2022-QPT}.
Our results in Table \ref{Tab:Rad} are up to $60$ times more accurate than those obtained with a CKP method \cite{2004-FrickeCd}.

\section{Towards BSM physics}

We can use the above results to assess the requirements for King Plots in Cd I ISs to be sensitive to new physics.
An interaction between electrons and neutrons that is mediated by a heavy boson has a short range, less than a Bohr radius. To leading order, it is a contact interaction, proportional to the wave function overlap with the nucleus, and therefore proportional to the FS.
Such a contribution is absorbed into charge radii differences, so it does not create deviations of linearity in KPs.
In the next order, and in a hydrogenic approximation, the deviation from linearity turns out to be proportional to the FS \cite{2017-KPnon}. Therefore, a KP for transitions $i$ and $j$ where $|F_i - F_j|$ is large, can have a high sensitivity to such new physics.
In Fig. \ref{fig:Levels}b we plot $|F_i-F_j|$ for important transition pairs, whose atomic constants are given in Table \ref{tab:Atom}.
As expected, large differences are between pairs that do not share the same $S$ state. However, narrow transitions couple the ground $S$ state with the $5p$\,$^3P_j$ states, enabling exceptional precisions for their measurements.
Focusing on the clock and intercombination line pair, a KP returns $F_{332,326}=1.008(13)$, with an uncertainty largely from the $468\,$nm measurement \cite{1981-WTM} and our $480\,$nm measurement.
Multiplying by $F_{326}=-4354(62)$ MHz\,fm$^{-2}$ from Table \ref{tab:Atom}, we find $|F_{332}-F_{326}|\leq100\,$MHz\,fm$^{-2}$.
For the $S-D$ transitions used in the KP of Ca II, $|F_{732}-F_{729}|=0.55(2)\,$MHz\,fm$^{-2}\,$ \cite{2020-Ca}, which is up to two orders of magnitude smaller.
We therefore surmise that when searching for new interactions mediated by a heavy boson, the sensitivity of a KP between the ISs of the clock and intercombination lines in Cd could be as much as two orders more sensitive than the current limit set by IS measurements in Ca II.
As the accuracy of measurements in the latter are of order $20\,$Hz, accuracies of at least few kHz would be required for Cd I.
Considering that the natural linewidth of the intercombination line is $69\,$kHz, such an accuracy should be straightforward with trapped samples. 

%% SM nonlinearities
It is difficult to attribute an observed KP non-linearity to new physics without estimating non-linear SM contributions, such as the quadratic mass shift (QMS). 
Due to the similarity between nuclear masses and mass numbers, the deviation pattern of the QMS is nearly indistinguishable from the pattern induced by a new boson that couples electrons and neutrons, and so it sets a stringent limit on the new physics reach.
The order of magnitude of the QMS can be estimated with a hydrogenic approximation to be $K_{ij}^\textrm{QMS}=K_{ij}(m/M_A+m/M_{A'})$ \cite{2020-Ar},
with $|K_{332,326}|\leq60\,$GHz u from our KP.
A linear fit of $K_{332,326}^\textrm{QMS}$ versus $\delta \bar{\nu}_{326}^{A,A+2}$ returns an offset of $0.7\,$MHz u, a slope of $6\times10^{-10}$, and, most importantly, isotope-dependent non-linearities.
The largest deviation from the linear fit is $4\,$Hz, for the $^{106,108}$Cd pair and the maximum value of $K_{332,326}$. It is comparable to the estimated $3\,$Hz QMS-induced non-linearity in Ca II \cite{2018-NonlinCharge}.

% Mass uncertainty seems to have negligible effect on the deviation

%% QFS
For the medium mass Cd system, with its small nuclear deformations, the largest expected SM non-linearity is the quadratic field shift (QFS) \cite{2021-NP}.
In the notation of Eq. \ref{KP}, it takes the form $F_{ij}^\textrm{QFS}=G_{ij}(\delta(r^2)^{AA'})^2/\mu_{AA'}$, where $G_{ij}=G_i-F_{ij}G_j$. $G_{ij}$ sets the scale of the non-linearity and is a transition-dependent and nucleus-independent constant \cite{2021-Yb,2020-Yb}.
This scale is difficult to estimate and different calculations yield a wide range of sensitivities \cite{2020-Yb}.
Assuming that the QFS is approximately quadratic in $|F_i-F_j|$, the absence of observed KP non-linearities in this work already indicates that the non-linearity induced by the QFS of the narrow transition pairs is less than $1\,$kHz.
Moreover, the QFS may be identified from the deviation pattern from linearity \cite{2021-Yb}. The precisions with which $\delta(r^2)$ are known set the limits for this identification, thus emphasizing the need for the improved radii in Table \ref{Tab:Rad}.

%% GKP
Other sources of non-linearity, such as nuclear polarization and nuclear deformation, could be non-negligible at the anticipated level of precision  \cite{2021-NP,2021-NP2,2021-NPYb}. Calculating their magnitude and pattern is difficult due to our limited understanding of nuclear structure. However, like the QFS, their magnitude increases with the wave function overlap with the nucleus.
By combining the ISs of three or more transitions with different FSs, these non-linearities can be isolated  \cite{2017-MKP, 2020-GKP}.
Based on the above considerations, the ISs of the clock and intercombination lines can be measured to the order of $\sim100\,$Hz, limited by the natural linewidth of the intercombination line.
If no non-linearity is observed, the ultra-narrow $314\,$nm line may provide higher precision with reasonably controlled tensor light shift.
If a non-linearity is present at this level, the IS of a transition that does not include the ground state can lead to a sensitive multi-dimensional KP.
As suggested by Fig. \ref{fig:Levels}b, the $|F_i-F_j|$ for these transitions are at least two orders of magnitude larger than for ground state transitions.
Narrow transitions between excited states, such as from $^{3}P_{0}$ to long-lived low-lying Rydberg states, can thus increase the sensitivity of searches for BSM physics.

\section{Summary}

The  measurements reported here significantly improve the isotope shifts of several low-lying transitions of Cd I, including the first measurements of two transitions. 
For the wide $229\,$nm and the narrow $326\,$nm transitions, we utilize a cryogenic beam with enriched samples, combined with a detection method that distinguishes between the emission patterns of bosonic and fermionic isotopes.
We also measured ISs of the $326$ and $361\,$nm transitions using the sharp blue edge of laser-cooling in a magneto-optical trap, and of the $480\,$nm transition via  optical pumping of the trapped atoms. %for the $480\,$nm transition through the effect of optical pumping on the MOT population.
Combining our $480$ and $326\,$nm ISs with previous ISs for the $486\,$nm \cite{1981-WTM} and $509\,$nm \cite{2018-ISOLDE} transitions, we predict the ISs of the two ultra-narrow transitions in Cd, which have not yet been measured. 
Our results are significantly more accurate, often in disagreement with previous measurements, and highlight the benefits of isotope-selective and cold sources for IS measurements.

We also present high-precision calculations of IS constants of Cd II.
By projecting our calculations from Cd II to Cd I, using King Plots with measured ISs, we find the IS constants for all low-lying transitions in Cd I, including those of the narrow intercombination and ultra-narrow lines. 
Our resulting IS constants largely agree with, and are more accurate than those obtained from the CI-MBPT \cite{2021-Julian} and MCDHF methods \cite{2021-SM}, setting stringent benchmarks for improving the theory. 
Some disagreements are observed with recent calculations \cite{2021-SM}, potentially from underestimating the uncertainty of high-order electron correlations as well as the importance of the QED contributions to the field shift constants.

By combining our calculated IS constants with measurements of a long chain of short-lived ions, we extract accurate charge radii differences, without the limitations from muonic atom measurements.
To our knowledge, this is the first extraction of charge radii differences for a chain of isotopes with an accuracy better than $1\%$, which opens opportunities to improve our knowledge of nuclear sizes far from stability.
Beyond benchmarking nuclear models, these are important to identify the patterns of non-linearities in future high-precision measurements \cite{2021-Radii}.
This work sets the stage for new physics searches using generalized KPs in the Cd system.
Our results suggest that precise future Cd IS measurements can improve, by as much as two orders of magnitude, the current best bounds on new electron-neutron interactions, obtained from a KP for Ca II transitions. 

\section*{Acknowledgement}
We thank Ronnie Kosloff for stimulating conversations, Eberhard Tiemann for helpful assistance, and Yotam Soreq for valuable suggestions. 
We gratefully acknowledge the use of the Vikram-100 HPC cluster of Physical Research Laboratory, Ahmedabad (B.K.S.) and financial support from the US National Science Foundation (K.G.) and the European Research Council (ERC) under the European Union’s Horizon 2020 Research and Innovation Programme (CoMoFun, S. T.).

\bibliography{sample.bib}

\end{document}